\documentclass[twocolumn,amssymb,prb,showpacs]{revtex4}
\usepackage{graphicx}
\usepackage{dcolumn}
\usepackage{bm}

\begin{document}

\title{Local Density of States in Mesoscopic Samples from Scanning Gate Microscopy}

\author{M. G. Pala,$^1$ B. Hackens,$^2$ F. Martins,$^3$
H. Sellier,$^3$  V. Bayot,$^2$ S. Huant,$^3$ and T. Ouisse$^3$}

\affiliation{$^1$IMEP-LAHC-MINATEC (UMR CNRS/INPG/UJF 5130), BP 257, 38016 Grenoble, France}

\affiliation{$^2$CERMIN, DICE Lab, UCL, B-1348 Louvain-la-Neuve, Belgium}

\affiliation{$^3$Institut N\'eel, CNRS, and Universit\'e Joseph Fourier, BP 166, 38042 Grenoble, France}

\begin{abstract}
We study the relationship between the local density of states (LDOS) and the conductance  
variation $\Delta G$ in scanning-gate-microscopy experiments on mesoscopic structures
as a charged tip scans above the sample surface. We present an analytical model	
showing that in the linear-response regime the conductance shift $\Delta G$ is 
proportional to the Hilbert transform of the LDOS
and hence a generalized Kramers-Kronig relation 
holds between LDOS and $\Delta G$.
We analyze the physical conditions for the validity of this relationship
both for one-dimensional and two-dimensional systems when several 
channels contribute to the transport.
We focus on realistic Aharonov-Bohm rings including a random distribution of impurities and analyze
the LDOS-$\Delta G$ correspondence by means of exact numerical simulations,
when localized states or semi-classical orbits characterize the wavefunction of the system.
\end{abstract}


\date{\today}

\maketitle

\section{Introduction}
\label{s1}

Scanning probe microscopy (SPM) is nowadays an essential technique 
to measure local electronic properties of mesoscopic structures.
Scanning tunnel microscopy (STM), consisting in 
probing the sample surface with a metallic tip, is the most popular among all SPM techniques. 
STM experiments have first enabled
the detection of standing wave pattern in confined surface electron systems such as
quantum corrals.\cite{cromnie} The physical interpretation
of such images is immediate since they derive from direct tunneling 
between the surface electrons and the tip.
Hence, STM images represent the density of states of the system at a given 
position of the scanning probe.\cite{cromnie}

Later on, another SPM technique, named
scanning gate microscopy (SGM), has been introduced in order to obtain similar information for
structures buried under an insulating layer.
This technique consists in scanning a biased tip over the sample surface.
The gate acts as a local electrostatic (repulsive or attractive) potential 
on the electronic system and allows to obtain
two-dimensional (2D) conductance (or resistance) images of the scanned area 
as a function of the tip position.
At the present time, SGM or an alternative technique
called scanning capacitance microscopy (SCM) have been adopted to investigate the physics of
quantum points contacts,\cite{topinka1, topinka2, aoki1, decunha, jura}
quantum dots,\cite{pioda, fallahi} carbon nanotubes,\cite{bachtold} open billiards\cite{crook}
and edge states in the integer quantum Hall regime.\cite{finkelstein, aoki2, baumgartner, weis}
SGM on InAs nanowires has evidenced the presence
of multiple quantum dots inside the structure corresponding to
circular Coulomb blockade peaks in the conductance plots.\cite{blesznynski}

From a theoretical point of view,
numerical simulations based on the Keldysh-Green's functions formalism
have been exploited to demonstrate wavefunction imaging in open quantum dots\cite{mendoza1,mendoza2}
and to associate conductance images to
electron flow in quantum point contacts.\cite{metalidis,cresti}

Recently, isophase lines for electrons in an electrostatic 
Aharonov-Bohm (AB) experiment\cite{hackens1}
and local-density-of-states (LDOS) mapping inside a coherent AB ring\cite{martins} 
have been reported.
In the latter work, both experimental curves 
and numerical simulations have found a linear dependence of the conductance
variation on the tip voltage.
Interestingly, clear wavefunction images were obtained only 
in this case,\cite{martins} suggesting to limit the imaging technique to 
the linear-response regime.
It was also noticed that the simulated conductance maps were not significantly
influenced by the particular shape adopted to mimic
the effective perturbing potential due to the scanning gate.

Since SGM is essentially an indirect measurement
of the 2D electron gas (2DEG) properties, a physical interpretation
of conductance images is not immediate.
Therefore, in this paper we try to clarify the physical 
meaning of SGM experiments and isolate the
experimental conditions under which the linear response-regime holds.

In Section \ref{s2} we present an analytical model 
which provides a simple physical interpretation of the SGM conductance images
in the single-channel transmission regime.
We find that the conductance correction due to the scanning tip
is related to the Hilbert transform of the local density of states (LDOS) of the system.
Moreover, we analyze the spatial and energetic conditions on the 
perturbing potential under which this direct relation is maintained.

In Section \ref{s3} we present 2D numerical simulations of a 
realistic quantum ring characterized by the
presence of randomly localized impurities. 
Our calculations are based on a recursive Green's functions method 
and illustrate the correspondence
between LDOS and conductance images of the system in such a complex case.
In particular, we address the interesting case for which the wavefunction of the system
is dominated by localized states due to charged impurities
or by recursive semi-classical orbits.

Section \ref{s4} summarizes the relevant results.

\section{Single-channel transmission}
\label{s2}
In this Section, we evaluate the effect of a local perturbation 
on the total transmission of a non-interacting system
connected to single-channel leads.
We adopt the Landauer-B\"uttiker transport
theory assuming the zero--temperature limit.\cite{landauer} 

We consider a multi-states system connected to one-dimensional (1D) leads
and neglect the effect of inelastic scattering and electron-electron interactions.
These assumptions are valid in
the case of low-temperature experiments on mesoscopic samples
working in the phase coherent transport regime.\cite{hackens2}
We model our system as a chain of $N$ sites with the $i$-th on-site potential $\epsilon_i$
coupled to two semi-infinite reservoirs with fixed chemical potentials.
The coupling strength with the left and the right reservoir is denoted by 
$\Gamma_{\rm L,R}=2\pi|V_{\rm L,R}|^2 \rho_{\rm L,R}$,
where $V_{\rm L,R}$ is the coupling with the leads and $\rho_{\rm L,R}$
is the density of states of the reservoirs, respectively.
Moreover, each site is coupled only to its nearest neighbor with coupling constant $V$,
which for simplicity we assume not to depend on the position.

Since the linear conductance of a system with Fermi energy $\epsilon_{\rm F}$ 
is determined by the propagator from site 1 to site $N$,
we simply need to evaluate the element $G^{\rm R} (x_N,x_1)=G^{\rm R}_{N,1}$
of the retarded Green's function of the system $G^{\rm R}$.
By adopting the common assumption of a large band-width coupling with the reservoirs,
the conductance $G$ reads
\begin{equation}
G(\epsilon_{\rm F})=G_0 \Gamma_{\rm L}\Gamma_{\rm R}\left| G^{\rm R}_{N,1} (\epsilon_{\rm F}) \right|^2 ,
\label{trans} 
\end{equation}
where $G_0$ is the quantum of conductance.\cite{datta}
Further, we will use the well known fact that the LDOS at $x_i$ is proportional to
the imaginary part of the diagonal elements of the retarded Green's function $G^{\rm R} (x_i,x_i)=G^{\rm R}_{i,i}$
as 
\begin{equation}
\rho (x_i; \epsilon_{\rm F})=-\frac{1}{\pi} {\rm Im} \left[G^{\rm R}_{i,i} (\epsilon_{\rm F})\right],
\label{ldos}
\end{equation}
which derives from the pole structure of the Green's function.\cite{mahan}
In order to relate the conductance in Eq.~(\ref{trans}) and the LDOS in Eq.~(\ref{ldos}) we exploit
the Dyson equation
\begin{equation}
G^{\rm R}=g^{\rm R}+g^{\rm R} \hat{V} G^{\rm R}=g^{\rm R}+G^{\rm R} \hat{V} g^{\rm R} ,
\label{dyson}
\end{equation}
where $g^{\rm R}$ is the unperturbed Green's function of the isolated sites
and $\hat{V}$ corresponds to the hopping connection between sites.
From Eq.~(\ref{dyson})
one can obtain the relation between the propagator $G^{\rm R}_{i+1,1}$ 
and $G^{\rm R}_{i,i}$  or  $G^{\rm R}_{i+1,i+1}$ as\cite{sols} 
\begin{equation}
G^{\rm R}_{i+1,1}= G^{\rm R}_{i+1,i+1} V g^{\rm R}_{i,i}=g^{\rm R}_{i+1,i+1} V G^{\rm R}_{i,i}.
\label{dyson2}
\end{equation}
Hence, by recursively applying Eq.~(\ref{dyson2}) to express $G^{\rm R}_{N,1}$ in terms of $G^{\rm R}_{i,i}$, 
we can rewrite Eq.~(\ref{trans}) as
\begin{equation}
G=G_0 \Gamma_{\rm L}\Gamma_{\rm R} |V|^{2(N-1)} \left| G^{\rm R}_{i,i} \right|^2  \prod_{j \ne i} \left| g^{\rm R}_{j,j} \right|^2  ,
\end{equation}
where the index $i$ can have any value between 1 and $N$.
Further, since we are interested in studying the stationary regime of transport,
it is convenient to write the Green's functions in the energy space representation 
by performing a Fourier transform.
We have
$g^{\rm R}_{i,i}=[\epsilon_{\rm F}-\epsilon_i+ i 0^+]^{-1}$ and								
$G^{\rm R}_{i,i}=[(g^{\rm R}_{i,i})^{-1}-\Sigma_i]^{-1}=[(g^{\rm R}_{i,i})^{-1}-\Lambda_i+i\Gamma_i]^{-1}$,
where $0^+$ is a positive infinitesimal number,
$\Sigma_i=(\Lambda_i-i\Gamma_i)$ is the self-energy including the interactions with the system 
on the local $i-$th Green's function.\cite{mahan}

We now consider the effect on the transmission of a local perturbing potential which can be 
physically associated with the action of a charged tip scanning the system. 
We mimic the effect of the charged tip at position $x_i$
as a renormalization of the on-site energy $\epsilon_i +\Lambda_i \rightarrow \epsilon_i +\Lambda_i + U$,
where $U$ is the perturbing potential.
Assuming that the potential strength is sufficiently ``weak'',
the conductance correction is linear in $U$
and reads
\begin{equation}
\frac{\Delta G}{G} (x_i; \epsilon_{\rm F})= 2 U {\rm Re}  \left[G_{i,i}^{\rm R} (\epsilon_{\rm F}) \right] .
\label{corr}
\end{equation}

By comparing Eq.~(\ref{corr}) with Eq.~(\ref{ldos}) we find that
the conductance variation due to the action
of a local perturbing potential and the LDOS are related in the same way as 
the real and the imaginary part of the Green's function $G_{i,i}^{\rm R}$,
namely by a Kramers-Kronig relation.
Mathematically, this is expressed in terms of the Hilbert transform
${\rm Re} \left[ G_{i,i}^{\rm R} (\omega) \right]= {\cal H} \{ {\rm Im} \left[ G_{i,i}^{\rm R} (\omega) \right ]\}
=\frac{1}{\pi} {\cal P} \int d\omega' \frac{1}{\omega-\omega'}{\rm Im} \left[ G_{i,i}^{\rm R}(\omega')\right]$,
where ${\cal P}$ stands for the Cauchy principal part.
Therefore, in the linear-response regime, 
the conductance (transmission) correction
is proportional to the Hilbert transform of the local density of states. 
Notice that the sign of the conductance correction
is determined by both the perturbing potential $U$
and by the relative position of the Fermi energy with respect to $\epsilon_i+\Lambda_i$.

The physical origin of such a relationship can be understood by noticing
that the main effect of the local perturbing potential $U$
is to renormalize the real part of the self-energy.

By analyzing Eq.~(\ref{corr}) we can deduce the conditions
of validity of the LDOS-$\Delta G$ correspondence.
In order to deal with a ``weak'' perturbation 
giving rise to a linear correction the condition $|U|<|\epsilon_{\rm F}-\epsilon_i-\Lambda_i|$
has to hold.
A condition on the extension of the perturbing potential
is given by the spatial periodicity of the wavefunction
and limits the spatial range
of the effective potential
below the half-Fermi wavelength $\lambda_{\rm F}/2$.
Nevertheless, notice that in the multichannel transmission case
that we discuss in Section \ref{s3},
the spatial periodicity of the LDOS can be larger than $\lambda_{\rm F}/2$.

Interestingly, in the single-channel transmission regime
the LDOS is usually a sinusoidal function
of the spatial coordinate, 
whereas the Hilbert transform
of such a function alters the original function by only a $\pi/2$ phase shift.
More precisely, we have that the Hilbert transform adds a $+\pi/2$ phase
for right-going states and $-\pi/2$ for left-going states.
This justifies the use of a scanning local perturbation, 
like that obtained with SGM experiments, 
as an efficient tool to image the LDOS of the system.

\subsection{Symmetric double-barrier}

\begin{figure}[h!]
\begin{center}
\includegraphics[width=3.2in]{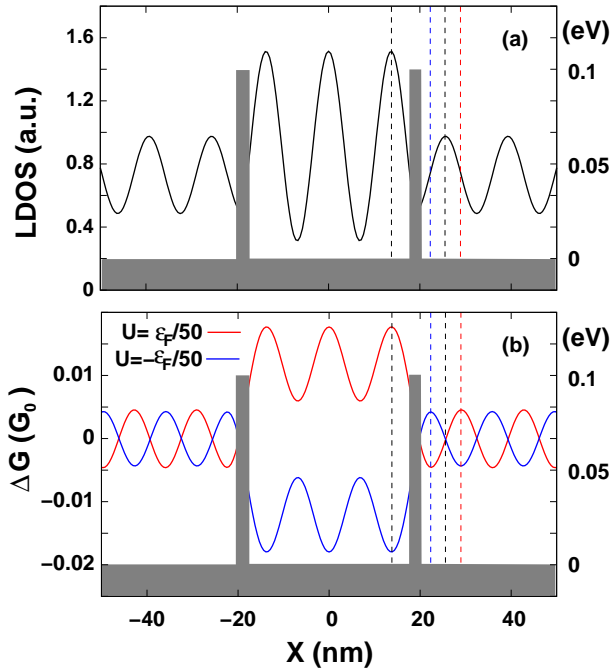}
\caption{(a) LDOS and (b) conductance variation $\Delta G$  
for a double-barrier system. The distance between the two barriers of height 0.1 eV is $L=18$ nm, 
the effective mass is $m=0.04 m_0$ with $m_0$ the free electron mass 
and the Fermi energy is $\epsilon_{\rm F}=0.05$ eV.
The perturbing potential used to obtain (b) is a square repulsive (red line) or attractive (blue line) 
barrier with height $U_0=\pm \epsilon_{\rm F}/50$ and
width $W=10$ nm. Dashed vertical lines indicate the maximum position of LDOS (black line)
and $\Delta G$ (red or blue line).}
\label{figure1}
\end{center}
\end{figure}

In order to give a practical application of the model we study 
the simple case of a symmetric double barrier.
In Fig.~\ref{figure1} we show the LDOS of the unperturbed system 
and the conductance correction due to a scanning potential 
acting as a local scatterer. 
The conductance is computed by means of Eq.~(\ref{trans})
after numerical calculation of the total Green's function. 
The local perturbing potential is assumed to be a rectangular-shaped
barrier with height $U_0=\epsilon_{\rm F}/50$ (or $U_0=-\epsilon_{\rm F}/50$)
and width $W=10$ nm smaller than $\lambda_{\rm F}/2 \sim 14$ nm.

Let us discuss the $\Delta G$-LDOS correspondence separately for the two regions 
inside and outside the two barriers.
We model the two barriers by using the same scattering matrix
$S$ which can be parametrized as
\begin{equation}
S=\left( 
\begin{array}{cc}
r & t \\ t & r 
\end{array}
\right) =\left( 
\begin{array}{cc}
e^{i\alpha} |r| &  i e^{i\beta} |t| \\  i e^{-i\beta} |t| & e^{-i\alpha} |r| 
\end{array}
\right) ,
\end{equation}
where $\alpha$ and $\beta$ are real phases ($\beta=0$ if
time-reversal symmetry holds) and $|t|=\sqrt{T}$, $|r|=\sqrt{1-T}$,
with $T$ transmission probability.
In order to compute the LDOS inside the two barriers
we evaluate the wavefunction for left- and right-injected states by
imposing the proper boundary conditions.
We obtain $\Psi_{\rm L}(x)\propto [e^{ik_{\rm F}(x-L/2)}+|r|e^{i\alpha} e^{-ik_{\rm F}(x-L/2)}]$
and  $\Psi_{\rm R}(x)\propto [e^{-ik_{\rm F}(x+L/2)}+|r|e^{-i\alpha} e^{ik_{\rm F}(x+L/2)}]$,
with $k_{\rm F}$ the Fermi wavevector and $L$ the distance between the two barriers.
By estimating the LDOS of the system as the independent sum of 
left-injected and right-injected states
we obtain $\rho(x)\propto[1+|r|^2+2|r|\cos(k_{\rm F}L) \cos(2k_{\rm F}x-\alpha)]$.
The Hilbert transform of this function
adds a $+\pi/2$ phase shift to the partial LDOS 
corresponding to the left-injected states and 
a $-\pi/2$ phase shift to the partial LDOS 
corresponding to the right-injected states.  
It reads ${\cal H} \{ \rho(x) \} \propto [1+|r|^2+2|r| \cos(k_{\rm F}L-\pi/2) \cos(2k_{\rm F}x-\alpha)]$
and it has the same spatial dependence of $\rho(x)$.
An identical result is numerically recovered in Fig.~\ref{figure1} 
by computing the conductance variation due to a square perturbing potential.
Notice that a further sign on the conductance variation can be 
determined by the choice of an attractive or repulsive potential.

The effect of the local scatterer is different in the spatial region outside the two barriers.
On the left (right) of the barriers the LDOS shape is determined
only by left (right)-injected states which gives 
$\rho(x) \propto {\rm const}+\cos(2k_{\rm F}(x \mp L/2)-\alpha)$. 
In fact on the left (right) of the barrier the right (left)-injected states contribute as a constant
to the LDOS.
The Hilbert transform of such a function gives the same expression with a $\pm \pi/2$ phase 
shift for right- and -left-injected states.
Again, this is the result found numerically  
in Fig.~\ref{figure1}.

Finally, we address the case the Fermi energy equals $\epsilon_i+\Lambda_i$ it turns out that
${\rm Re}  \left[G_{i,i}^{\rm R}) \right]=0$
and the system is ``on resonance''.
This condition gives a maximum of the transmission probability.
In such a case Eq.~(\ref{corr}) vanishes and
the correction to the transmission is given by the second order in $U$. It reads
\begin{equation}
\frac{\Delta G}{G} (x_i; \epsilon_{\rm F})
=-\pi^2 \rho^2 (x_i; \epsilon_{\rm F})  U^2.
\label{res}
\end{equation}

Therefore, in the resonant case 
the correction to the transmission is a monotonic function
of the local density of states $\rho(x_i)$ and has a fixed sign.


\subsection{Mesoscopic ring}

We focus now on the simulation of an AB ring since such a structure
is particularly suited to study phase coherence effects in mesoscopic physics.\cite{pala1}
Further, we consider a ring obtained from an InGaAs/InAlAs
heterostructure\cite{hackens2} with $W=120$ nm for
the width openings and $R_{\rm in}=140$ nm and  $R_{\rm out}=265$ nm 
for the inner and outer radius, respectively.
The effective mass of carriers is assumed to be $m=0.04 m_0$ with $m_0$ the free electron mass. 
For the present case we analyze an ideal ring without the presence of any defect.
We adopt a tight-binding model to describe the Hamiltonian of the
system.\cite{mireles}
The numerical technique is based on the calculation of the retarded
Green's function of the system by using a recursive technique.\cite{baranger,ferry,frustaglia,pala2}
The method is quite general and allows to treat the possible influence of spin-orbit 
coupling on the system.\cite{schapers}
The presence of an external magnetic field can be 
modeled by adding Peierl's phases to the
hopping elements of the Hamiltonian.\cite{pala2}

\begin{figure}[h!]
\begin{center}
\includegraphics[width=1\columnwidth]{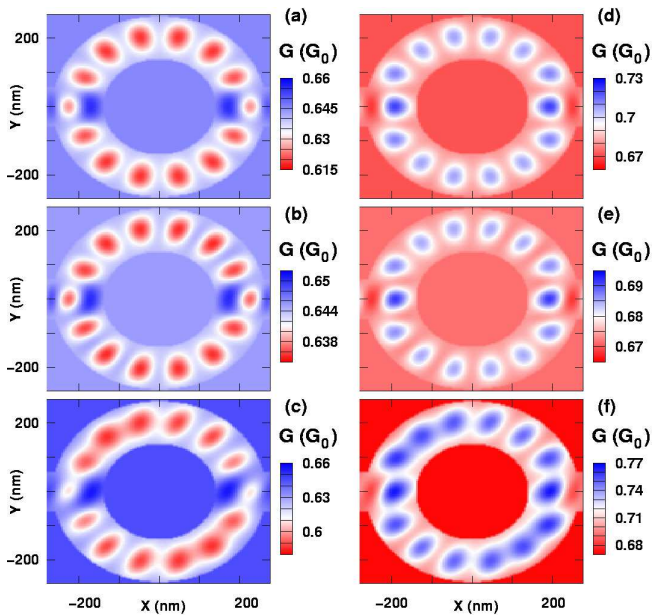}
\caption{Conductance maps for an ideal AB ring with $W=120$ nm for
the width openings and $R_{\rm in}=140$ nm and  $R_{\rm out}=265$ nm
for the inner and the outer radius, respectively.
We have adopted a Fermi energy $\epsilon_{\rm F}=1.65$ meV in (a), (b), (c),
and $\epsilon_{\rm F}=1.75$ meV in (d), (e), (f), 
whereas a maximum for the transmission energy stays at $1.7$ meV.
Therefore, figures on the left stay below and
figures on the right stay above a resonance.
For the first-line couple (a), (d) the perturbing potential mimicking the effect of the charged tip
is modeled as $U(r)=U_0/\{1+[r/\sigma]^2\}$ with $U_0=\epsilon_{\rm F}/10$ and $\sigma=10$ nm.
For the couple (b), (e) and the couple (c), (f) we have adopted the same asymmetric perturbing 
potential $U(r)=U_0/\{1+[(x+y)/\sigma]^2+[2(x-y)/\sigma]^2\}$
with $U_0=\epsilon_{\rm F}/10$,
but with different spatial extension determined by $\sigma=10$ nm and $\sigma=40$ nm, respectively.}
\label{figure2}
\end{center}
\end{figure}

We study the conductance variation
of such an AB ring under the effect of a superimposed perturbing potential $U(x,y)$.
In Figs.~\ref{figure2}(a),  \ref{figure2}(d) we have  
first assumed a long-range Lorentzian function $U(r)=U_0/\{1+[r/\sigma]^2\}$,
where $r=\sqrt{x^2+y^2}$, the coordinates $x$ and $y$
have the tip position as origin, $U_0$ is the maximum potential perturbation and $\sigma$
determines the decay rate of the potential. Notice that the 
spatial extension of such a potential is larger than $2\sigma$.
However, since recent experimental results on 
open quantum dots have pointed out that, in some cases, a significant spatial asymmetry characterizes
the tip-induced potential,\cite{gildemmeister}
we have also adopted an asymmetric shape for  $U(x,y)$.
In Figs.~\ref{figure2}(b),  \ref{figure2}(e) and Figs.~\ref{figure2}(c),  \ref{figure2}(f) we have assumed 
the potential $U(x,y)=U_0/\{1+[(x+y)/\sigma]^2+[2(x-y)/\sigma]^2\}$
with $\sigma=10$ nm and $\sigma=40$ nm, respectively.
Conductance maps on the left side of Fig.~\ref{figure2} have a Fermi energy
slightly smaller than a resonance point for which the ring has maximal transmission
$T=1$, whereas on the right side they have a Fermi energy slightly larger than the resonance.

Since both energies are close each other, the mapped conductances are very similar
and differ only for having opposite signs in complete agreement with Eq.~(\ref{corr}).
Here, we want to emphasize that, even if the 
correct potential shape of the perturbing potential 
should be obtained by computing the electrostatics of the total system, 
still the conductance variation does not depend on the particular
shape of the potential in the linear-response regime under study.\cite{martins}

Concerning the effect of an asymmetric tip-induced potential we observe
that the small asymmetry in Figs.~\ref{figure2}(b), \ref{figure2}(e)
can be considered negligible with respect to Figs.~\ref{figure2}(a), \ref{figure2}(b),
obtained with a symmetric potential.
On the contrary, it is relevant in Figs.~\ref{figure2}(c), \ref{figure2}(f) where the
spatial extension of the perturbing potential become larger
than $\lambda_{\rm F}/2 \sim 72$ nm giving rise to a non-linear behavior of the conductance.
We conclude that the effect of an asymmetric potential
can be minimized by providing a tip-induced perturbation with limited spatial extension.

\section{Multi-channel transmission}
\label{s3}
In this section we extend the results obtained for the single-channel transmission  
to the case the conductance is the sum of several independent 
channels as expressed by the Landauer-B\"uttiker formula
$G=G_0 \sum_n T_n$, where $T_n$ is the transmission probability of
the $n$-th conducting transverse mode.
This is the most common experimental situation. \cite{hackens1,martins}
In particular, we want to verify whenever the LDOS-$\Delta G$ correspondence
depicted in Section \ref{s2} is valid for realistic 2D structures.

A first complication arises from the spatial resolution of the 
scanning tip.
Indeed, in the  multi-channel transmission case
the LDOS structure is usually very complex 
and presents a spatial periodicity related to half the Fermi wavelength,
whereas in experiments the extension of the perturbing potential is typically larger than $\lambda_{\rm F}/2$.
Thus, one can expect that the spatial resolution
of the tip-induced potential is smaller than the spatial extension of 
conducting modes related to lowest energies, but larger than
those related to highest energies. 
This would lead to properly take into account
only conducting modes corresponding to the lowest energies
and lose a satisfactory correspondence with the LDOS.

Another problem concerns the interpretation of conductance maps 
in terms of Hilbert transform of the LDOS.
Generally, in the multi-channel transmission case it is not possible 
to obtain a direct correspondence like in Eq.~(\ref{corr}) or (\ref{res})
since the effect of the local perturbing potential due to the scanning tip
is different for each conducting channels contributing to the conductance.
However, it often occurs that the LDOS and hence the conductance are
dominated by one or few close transverse modes. 
In these specific cases the same LDOS-$\Delta G$ correspondence for the one-channel case is recovered
provided that the conditions found in Section \ref{s2} are valid.
This means that the spatial extension of the tip-induced potential
has to be smaller than the spatial periodicity of the dominant channel
{\it and}
the potential height $U_0$ has to be smaller than the corresponding energy level.

\begin{figure}[h!]
\begin{center}
\includegraphics[width=3.2in]{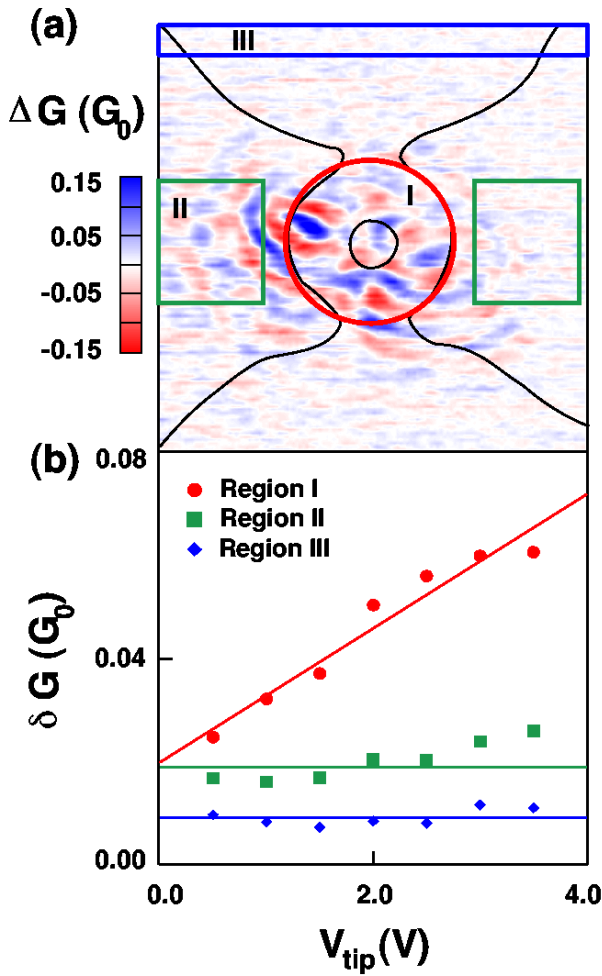}
\caption{
(a) Experimental conductance map measured for $V_{\rm tip}=$ 2.5 V 
and (b) averaged conductance variation $\delta G$
as a function of the tip voltage $V_{\rm tip}$
measured in different device regions indicated in (a).
A clear linear dependence is present only in the region I inside the ring.
The 2DEG distance of the tip is $D_{\rm tip}$= 50 nm, $T=4.4$ K and
a magnetic field of $B=2$ T is applied 
perpendicularly to the surface.}
\label{figure3}
\end{center}
\end{figure}

From an experimental point of view
we demonstrate the feasibility of the linear-response regime in Fig.~\ref{figure3}(b), 
where an averaged conductance variation as a function of the tip voltage is shown
for a coherent quantum ring.
This is fabricated from a InGaAs/InAlAs heterostructure using electron
beam lithography and wet etching. The 2DEG is located
25 nm below the surface, and its low temperature
($T$) electron density and mobility are $2 \times 10^{16}$ m$^{-2}$ and
10 m$^2$/Vs, respectively. The ring
inner and outer (lithographic) radii are 120 nm and 290 nm
(this sample is the same as sample R1 in Ref.~\onlinecite{martins}).
Notice that a clear linear dependence on the tip voltage is experimentally
found only if the tip scans the region I inside the ring,
where the former conditions are expected to hold. 

In summary, specific situations exist such that
the total wavefunction of the system adopts a simple and
regular shape and the conditions for a
satisfactory LDOS-$\Delta G$ correspondence are recovered.
In the following of this Section we give a few relevant examples 
and illustrate them with experimental data and numerical simulations.

\subsection{Localized states in AB rings with impurities}

\begin{figure}[h!]
\begin{center}
\includegraphics[width=3.2in]{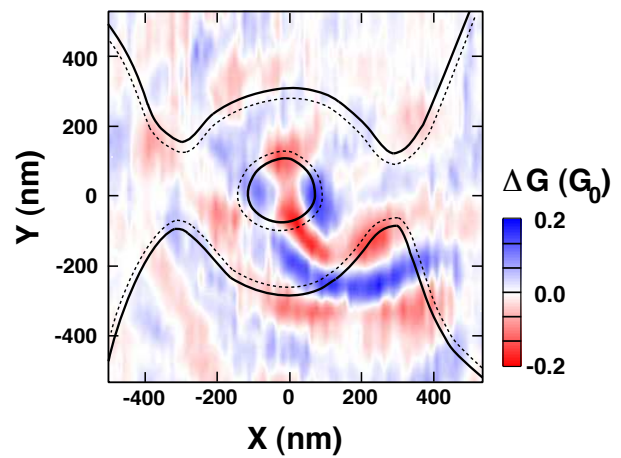}
\caption{
Experimental conductance map as a function of the tip position in the same sample as in Fig.~\ref{figure3}.
The dashed line indicates a depletion region at the border of the ring. 
$V_{\rm tip}$=0.5 V, $D_{\rm tip}$= 50 nm, $T=4.2$ K and $B=1.5$ T.
}
\label{figure4}
\end{center}
\end{figure}

We first focus on a realistic AB ring with the same sizes as the one
studied in Section \ref{s2}, but with a random distribution of symmetry-breaking defects. 
These defects are probably responsible for the 
asymmetric conductance images in Ref.~\onlinecite{martins},
where the wavefunction inside the ring is strongly influenced by localized states.
In Fig.~\ref{figure4} we display
a typical conductance map for the same ring of Fig.~\ref{figure3}.
We notice that the conductance map inside the ring is characterized by radial 
asymmetric fringes in both Fig.~\ref{figure3}(a) and Fig.~\ref{figure4}.

\begin{figure}[h!]
\begin{center}
\includegraphics[width=3.2in]{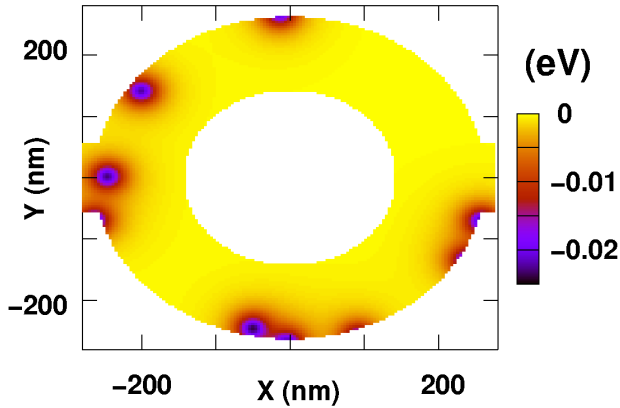}
\caption{Potential profile due to randomly distributed impurities
in a AB ring ($W=120$ nm, $R_{\rm in}=140$ nm and  $R_{\rm out}=265$ nm). 
These impurities are mainly localized around the border of the ring.}
\label{figure5}
\end{center}
\end{figure}

Here, we model the defects 
by superimposing a randomly distributed analytical potential $V(r)$ to the structure profile.
This takes into account the electrostatic effects of defects with charge $Q$ at distance $z$ from the 2DEG and reads
\begin{equation}
V(r) =\frac{Q}{4\pi \epsilon_0\epsilon_r\;z}\;
        \frac{I(a)}{1+a\;I(a)\left[\left(1+r^2/z^2\right)^{3/2}-1\right]} \, ,
\end{equation}
with
$a = \frac{e^2 m \;z}{2\pi\hbar^2\;\epsilon_0\epsilon_r}$
and $I(a) = \int_0^\infty{\frac{x\;e^{-x}}{x+a}\;dx}$.\cite{stern}
In our simulations we have adopted a random distribution
with a 90 $\mu$m$^{-2}$ numerical density at a fixed distance of 20 nm from the 2DEG. 
In Fig.~\ref{figure5} we show
the impurity-induced potential profile used in our calculations.

Further, we have chosen a Fermi energy range for which 
more than ten conducting channels are present in the openings 
and more than one hundred inside the ring.

The first favorable situation occurs
when the LDOS is determined by the position of these defects.
By computing the LDOS of the AB ring with the addition
of randomly distributed impurities we have found several images 
where the squared wavefunction of the system is dominated by
such localized states.

\begin{figure}[h!]
\begin{center}
\includegraphics[width=3.2in]{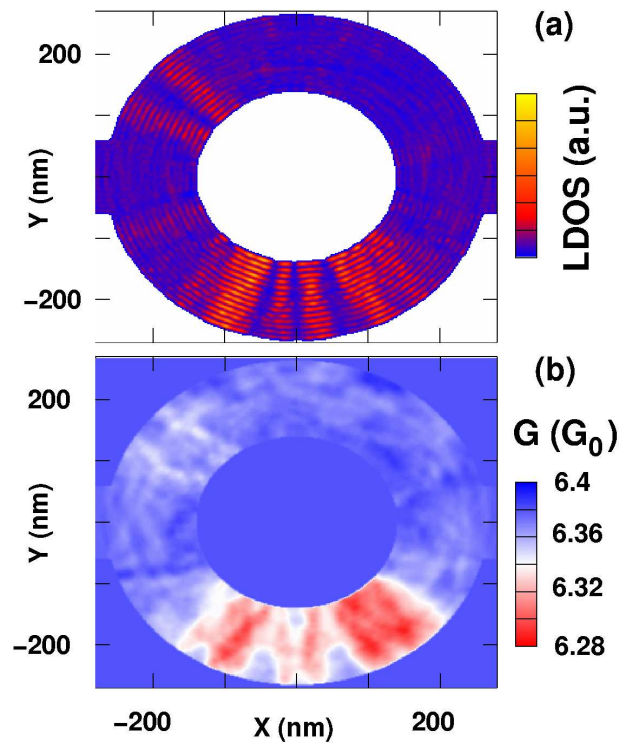}
\caption{(a) LDOS and (b) conductance map
of a realistic AB ring ($W=120$ nm, $R_{\rm in}=140$ nm and  $R_{\rm out}=265$ nm) 
with randomly distributed impurities. 
The presence of localized states is clearly visible in both of them.
The perturbing potential mimicking the effect of the charged tip
is modeled as the long range potential $U(r)=U_0/\{1+[r/\sigma]^2\}$ 
with $U_0=\epsilon_{\rm F}/100$ and $\sigma=5$ nm.
Other parameters are $\epsilon_{\rm F}=0.1074$ eV and $B$=0.}
\label{figure6}
\end{center}
\end{figure}

As an example, in Fig.~\ref{figure6}
we show the LDOS of the AB ring with the presence of randomly distributed impurities
for one of such specific energies,
as well as the corresponding conductance variation due to
a Lorentzian perturbing potential.
In this case the LDOS and the conductance variation are characterized by
radial asymmetric fringes similar to those of Fig.~\ref{figure4}.
Even if the effective spatial extension 
of the tip-induced potential is larger than $\lambda_{\rm F}/2 \sim 9$ nm
the conductance map accurately reproduces the
LDOS shape due to the large spatial extension of the localized states.
 
\subsection{Semi-classical orbits in AB ring with impurities}

Another favorable situation occurs in the presence
of semi-classical orbits inside the mesoscopic structure.\cite{akis, bird}
They correspond to quasi-classical states 
that scar the total wavefunction of the system and 
recur periodically as the Fermi energy or an
external magnetic field is varied.\cite{akis,bird}

\begin{figure}[ht!]
\begin{center}
\includegraphics[width=3.2in]{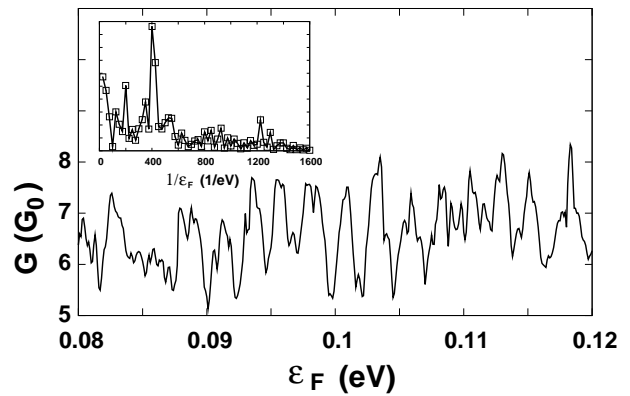}
\caption{Conductance as a function of the Fermi energy of the AB ring 
($W=120$ nm,$R_{\rm in}=140$ nm and  $R_{\rm out}=265$ nm) 
with randomly distributed impurities.
In the inset the Fourier transform is plotted.}
\label{figure7}
\end{center}
\end{figure}

\begin{figure}[ht!]
\begin{center}
\includegraphics[width=3.2in]{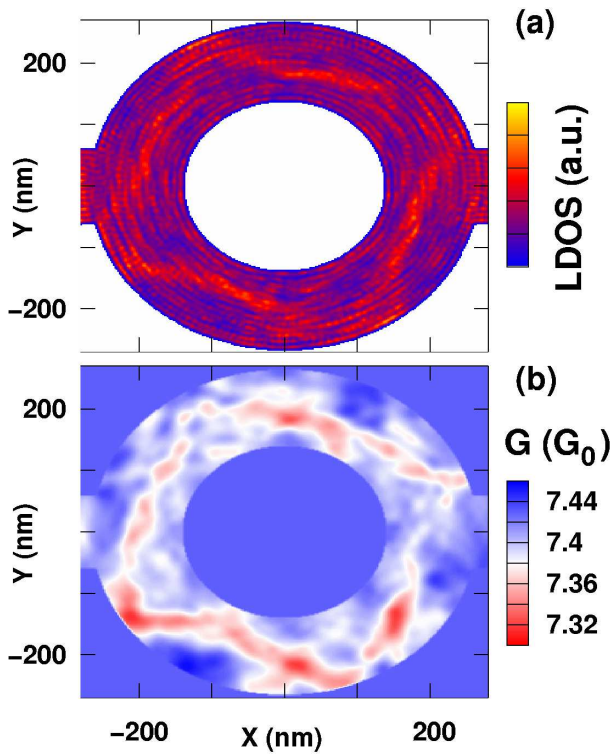}
\caption{(a) LDOS and (b) conductance map
of a realistic AB ring ($W=120$ nm, $R_{\rm in}=140$ nm and  $R_{\rm out}=265$ nm) 
with the presence of randomly distributed
impurities. Semi-classical orbits are well visible in both.
The perturbing potential mimicking the effect of the charged tip
is modeled as a the long range potential  
$U(r)=U_0/\{1+[r/\sigma]^2\}$ with $\epsilon_{\rm F}=0.1008$ eV, $U_0=\epsilon_{\rm F}/50$ and $\sigma=10$ nm.}
\label{figure8}
\end{center}
\end{figure}

In Fig.~\ref{figure7} we plot the conductance of the AB ring
with impurities as a function of the Fermi energy as well as its
Fourier transform (inset). In the latter plot it is possible to identify 
a sharp maximum at a frequency of 400 1/eV
corresponding to the occurrence of 
a dominant pattern in the LDOS.
Such a quasi-classical orbit is plotted in Fig.~\ref{figure8}(a)
and was found to recur with a frequency of 400 1/eV.
Also in this case the large and uniform spatial extension of the wavefunction
permits to recover a clear correspondence with the conductance map 
shown in Fig.~\ref{figure8}(b).

\section{Conclusion}
\label{s4}

We have presented a theoretical analysis of scanning gate microscopy 
as an experimental tool to image the LDOS of mesoscopic samples fabricated 
from buried 2D electron systems like Aharonov-Bohm rings.

We have shown that in the single-channel transmission case
a generalized Kramers-Kronig relation holds 
between the LDOS of the system and
the conductance variation due to the scanning tip.
Such relationship is valid in the linear-response regime 
where the tip induced potential is so weak 
that the energy levels of the system are unchanged.
In this case we have shown a striking correspondence between
LDOS and $\Delta G$ images for a symmetric double-barrier system
and an AB ring.

In the multi-channel transmission case an exact
LDOS-$\Delta G$ correspondence 
is less straightforward due to the deficient spatial resolution of the 
tip-induced potential even if the perturbing potential is smaller
than the energy levels of the system. 
However, in some special cases for which 
the total wavefunction presents a large and uniform spatial extension,
such correspondence can still be recovered.
In particular, we have shown a nice agreement between LDOS and conductance images
when the LDOS is dominated by localized states or by semi-classical periodic orbits
responsible for wavefunction scarring effects.

\begin{acknowledgments}
We would like to thank M. Governale for useful discussions and suggestions.
This work is supported by the {\it PNANO 2007} program of the 
{\it Agence Nationale de la Recherche} (``MICATEC'' project).
B. H. acknowledges funding by the FNRS (Belgium).
\end{acknowledgments}


\end{document}